\documentclass[
*aip,
reprint,
runinaddress,
showpacs,
preprintnumbers,
amsmath,amssymb,
aps,
pra,
floatfix,
]{revtex4-1}

\usepackage{mathtools}
\usepackage{float}
\usepackage{mathrsfs}
\usepackage{float}

\newcommand{\cm}{$\rm{cm^{-1}}$}

\begin{document}
\pagenumbering{arabic} 
\title{A versatile and narrow linewidth infra-red radiation source for ro-vibration state-selected preparation of molecules in  molecular beams}

\author{Avinash Kumar}%
\affiliation{Tata Institute of Fundamental Research Hyderabad, 36/P Gopanpally, Hyderabad 500046, Telangana, India}

\author{Saurabh Kumar Singh}%
\affiliation{Tata Institute of Fundamental Research Hyderabad, 36/P Gopanpally, Hyderabad 500046, Telangana, India}

\author{Pranav R. Shirhatti}
\email[Author to whom correspondence should be addressed. \\ e-mail:{\ }]{pranavrs@tifrh.res.in}
\affiliation{Tata Institute of Fundamental Research Hyderabad, 36/P Gopanpally, Hyderabad 500046, Telangana, India}%

\begin{abstract}
\textbf{Abstract:} 
We describe the design and characterization of a versatile pulsed (5 ns, 10 Hz repetition rate) optical parametric oscillator and amplifier system capable of generating single longitudinal mode, narrow linewidth (0.01 \cm{}) radiation in the wavelength range of 680 - 870 nm and 1380 - 4650 nm.
Using a combination of power-normalized photoacoustic signal and a Fizeau interferometer-based wavemeter, we are able to actively stabilize the output wavenumber to within 0.005 \cm{} (3$\sigma$) over a timescale longer than 1000 seconds.
We demonstrate an application of this system by performing ro-vibration state-selected preparation of CO in v = 2 state, via direct overtone excitation (v = 0 $\rightarrow 2$ at 2346 nm) and subsequent state-selected detection in an internally cold molecular beam.
\end{abstract}

\maketitle

\section{\label{sec:level1}Introduction}

Molecular beams are an essential tool for the preparation of molecules in a collision-less environment with well-defined momentum and internal state distributions, having high enough density for scattering experiments. 
The use of molecular beam and laser spectroscopy techniques, combined with surface preparation and characterization methods have enabled probing the dynamics of energy flow in chemical reactions at the gas-solid interface \cite{ Wodtke_chemical_2021,chadwick_quantum_state_2016,mccabe_molecularbeam_2000,bonn2002real}.
Molecule-surface interaction dynamics and energy transfer processes are important building blocks for understanding elementary steps in surface chemistry and heterogeneous catalysis.
Moreover, the well-defined conditions of the incident molecules and the surface in such experiments allow for a direct comparison with theoretical studies, making it valuable for benchmarking.

The collision-less environment and internal cooling in molecular beams result in a large fraction of molecules in the ground state with very narrow absorption linewidth for individual ro-vibrational transitions.
In the case where the beam of radiation used for excitation is perpendicular to the molecular beam (a common configuration for molecule-surface scattering experiments), these linewidths can be of the order of $10^{-4}$ \cm{}, mainly governed by the residual Doppler broadening. 
Therefore, for quantum state-selected preparation of molecules in such beams, high intensity, narrow linewidth, and frequency stable radiation source is required. 
The exact operating parameters, linewidth, and frequency stability, of a suitable radiation source also depend on the overall output power.
Narrow linewidth continuous (cw) radiation sources are particularly well-suited for such applications. 
For example, excitation of CH$_{4}$  in molecular beam environment has been successfully performed by several groups  \cite{Kliu_crossedbeam_2001,higgins2001state,chadwick2014quantum,juurlink1999eigenstate} in the context of state-selected scattering experiments.
Here, using radiation sources with linewidth less than $10^{-4}$ \cm{} and powers ranging up to hundreds of mW, efficient excitation has been demonstrated.
This scheme has also been extended for overtone transitions such as 2$\nu_{3}$ in CH$_{4}$ \cite{higgins2001state} and HCN \cite{srivastava2000sub} using specially designed resonant build-up cavities inside the vacuum chamber, to compensate for the low transition strength with high intracavity power. 

On the other hand, pulsed radiation sources also offer a possible route for state-selected molecular beam preparation. 
For example, excitation of HCl \cite{Wodtke_UHV_instrument_2007} and CH$_{4}$ \cite{RBeck_pulsed_dyelase_2003} have been demonstrated using difference frequency generation schemes with multimode dye-lasers.
However, the multimode nature of the radiation source leads to inefficient excitation.
Single longitudinal mode (SLM) pulsed injection seeded Optical Parametric Oscillators/Amplifier (OPO/A) systems overcome the above-mentioned limitation and have been used effectively for state-selected preparation in molecular beams \cite{Wodtke_inj_OPO_2010, wang_HIghpowerOPO_2020}. 
In these systems, the range of wavelength generated is limited typically to a few nanometers, mainly governed by the injection seed source, generally a wavelength-stabilized external cavity diode laser. 
Actively stabilized cw ring dye lasers and Titanium-Sapphire lasers can overcome this limitation but at the cost of added experimental complexity. 

Another interesting possibility is the use of a widely tunable single longitudinal mode, narrow linewidth master oscillator power amplifier (MOPA) design. 
This was demonstrated originally by Bosenberg and Guyer and later implemented by several other groups for high-resolution spectroscopy applications \cite{Deanguyer_broadlyTuunable_1993,Hogervorst_SLMOPO_2002,Zhang_pulsedSLMOPO_2017}.
Such systems in principle meet all the requirements for state-selected preparation of molecules in a molecular beam. 
Additionally, they also offer the versatility of being able to generate a very large range of wavelengths (typically 100 nm in visible and near-IR) without significant modifications.
However, only a few examples of the use of such sources for state-resolved scattering experiments are available in the literature \cite{korolik2000survival}. 
This is partly due to the fact that there are no ready commercially available alternatives at present. 
At the same time, wavelength stabilization and locking of low repetition rate, pulsed ns mid-IR radiation is generally challenging.

In this paper, we describe the design, construction, characterization, and demonstration of a versatile SLM-MOPA-based radiation source with active wavelength stabilization, suitable for molecular beam applications.
Using this source we perform CO (v = 0 $\rightarrow$ 2, 4263 \cm{}) direct overtone excitation in a molecular beam as an example for demonstrating the utility of the system for ro-vibration state-selected scattering experiments.
We also show that this design can be extended, with a few additional components, to generate near and mid-IR wavelengths to access CO$_{2}$ (v = 0 $\rightarrow$ 3, 6970 \cm{}, asymmetric stretch) and CO (v = 0 $\rightarrow$ 1, 2147 \cm{}) transitions.

\section{\label{sec:methods}Methods}

Our full experimental setup consists of a singly resonant OPO acting as a master oscillator based on a Littman-Metcalf cavity configuration \cite{littman_spectrallynarrowcavity_1978} and two single-pass amplifier systems, OPA-I  and OPA-II. All stages were pumped by the second harmonic of an injection seeded pulsed (5 ns full width at half maximum, FWHM) Nd$^{3+}$:YAG laser (Continuum Surelite III laser, 10 Hz repetition rate). The pump beam (532 nm, 7 mm diameter) is split into three parts to pump all three stages. For the Master oscillator, OPA-I and OPA-II, pump energy of 15 mJ/pulse, 65 mJ/pulse, and 80 mJ/pulse were used, respectively.
The pump beam diameter was adjusted using telescopes to 2 mm for OPO and 5 mm for OPA-I and OPA-II stages. 
\begin{figure*}
	\centering
	\includegraphics[width=1\textwidth]{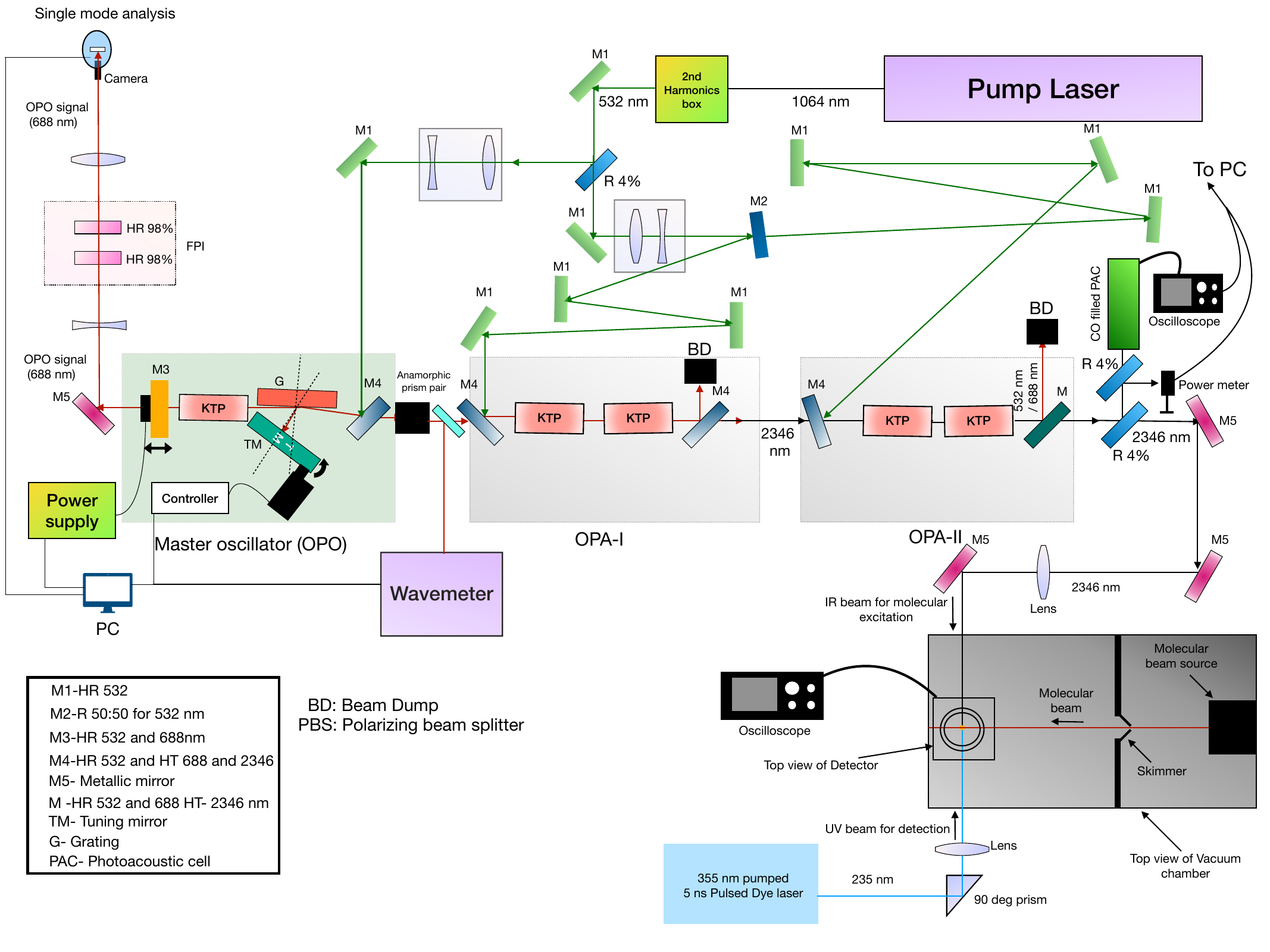}
	\caption{Schematic diagram of the full experimental setup including a master oscillator, OPA-I and OPA-II along with feedback control system, and IR-UV pump-probe experiment in the molecular beam.}
	\label{OPO_setup}
\end{figure*}
A schematic diagram of the master oscillator stage, referred to as OPO, is shown in Fig. \ref{OPO_setup}. 
The pump beam for the OPO was sent at a grazing incidence of 89 degrees on a reflective grating (holographic, 1800 lines/mm, 50 $\times$ 50 mm grating (G), GH50–18V, Thorlabs, Inc.). 
A type-II KTP (KTiOPO4) crystal 8 mm wide (in the phase matching plane)$\times$ 6 mm $\times$ 20 mm long (VM-TIM GmbH) with orientation $\theta$/$\phi$: 48.8/0 degree was used as a nonlinear gain medium (532 nm (o) = 688 nm (e) + 2346 nm (o)). A custom-coated cavity end mirror M3 (25 mm diameter, HR - 680 to 900 nm) was mounted on a piezo (PK44LA2P2, Thorlabs, Inc.) controlled translational stage (10 $\mu m$ displacement) to allow for cavity length control.
A 50 mm $ \times$ 50 mm protected silver mirror (PFSQ20-03-P01, Thorlabs, Inc.) was used for wavelength tuning (TM), by means of a piezo inertia motor controlled mount (PIAK10, Thorlabs, Inc.). 

The threshold pump energy for the oscillator cavity was around 8 mJ/pulse. With a pump energy of 15 mJ/pulse, it produced 200 $\mu$J/pulse of signal and idler combined.
The output of the OPO was taken from the zeroth order reflection of the grating.
The cavity optical length was kept short  (90 mm) for the largest possible free spectral range (FSR), enabling good mode discrimination.
The grazing angle of the grating was optimized for obtaining a low threshold and sufficient side-mode discrimination. 
SLM operation of the OPO was verified by sending a part of the OPO output into a  pair of mirrors acting as a Fabry-Perot interferometer (FPI) and analyzing the fringe pattern in real-time, captured using a webcam (see Fig. \ref{etalon}). 
This FPI consists of two plane mirrors placed 20.7 mm apart (R= 98$\%$ for 532 and OPO signal beam) resulting in an FSR of 7.2 GHz and a finesse of approximately 7.

The weak signal beam ($\sim$ 150 $\mu$J/pulse) generated from the master oscillator had an elliptical shape.
This was made into a circular shape by sending it through an anamorphic prism pair (PS877-B, Thorlabs, Inc.) and subsequently coupled to the OPA-I. 
A pair of KTP crystals, the same as that used in the master oscillator, were used in each amplifier stage. 
About $\sim$ 1 mJ/pulse of idler (2346 nm) power was observed at the OPA-I output and was sent into the OPA-II for further amplification. 
This resulted in $\sim$ 3-4 mJ/pulse of idler from OPA-II. 
This beam was focused using a 500 mm focal length CaF$_{2}$ lens for vibrational excitation of CO in a molecular beam. 
 
Angle tuning of the KTP crystal, placed on a manually controlled rotational stage (PR005/M, Thorlabs, Inc.), was used to coarsely change the output wavelength of the OPO. 
Using this method along with the in-line measurement of the OPO signal wavelength using a home-built wavemeter \cite{Wavemeter}, allowed us to reach within 1 \cm{} of the desired transition.
Thereafter, for day-to-day operation, we only needed to perform short SLM scans within the phase-matching bandwidth (3 to 5 \cm{}) of KTP crystal (held at a fixed angle) to reach the target wavelength. 
Automated SLM scans were performed with the help of a feedback system based on an in-line analysis of the fringe pattern observed using the FPI.

For active wavelength stabilization, some fraction of the idler output was sent into a CO-filled cell (20 mbar, 300 K, 500 mm long) for photoacoustic spectroscopy (PAS) measurements. 
PAS measurements in conjunction with the wavemeter (for signal wavelength) were used to determine the magnitude and direction of the wavelength drift in the mid-IR beam, respectively. 
This scheme allowed us to lock the idler wavelength for over $10^{3}$  second time scales routinely.
Vibrationally excited CO was prepared by IR excitation in a pulsed, skimmed molecular beam (nozzle: Parker 009-1643-900,  driver: IOTA ONE 060-0001-900, 10 Hz) and was detected using 2+1 resonantly enhanced multiphoton ionization (REMPI) scheme via the B$^{1}\Sigma^{+}$ state using $\sim$ 230 nm (UV) wavelength \cite{tjossem1989multiphoton}. 
The IR and UV beams were counter-propagating relative to each other with the molecular beam propagating perpendicular to both.
A home-built Wiley–McLaren time-of-flight mass spectrometer was used to detect the resonantly ionized molecules.
The wavelength measurement, feedback control for active wavelength stabilization, and SLM operation were implemented using a set of LabVIEW programs.

\section{\label{sec:res_disc}Results and discussion}

\subsection{\label{sec:OPO_performance}OPO performance characterization}

\begin{figure}
	\centering
	\includegraphics[width=0.40\textwidth]{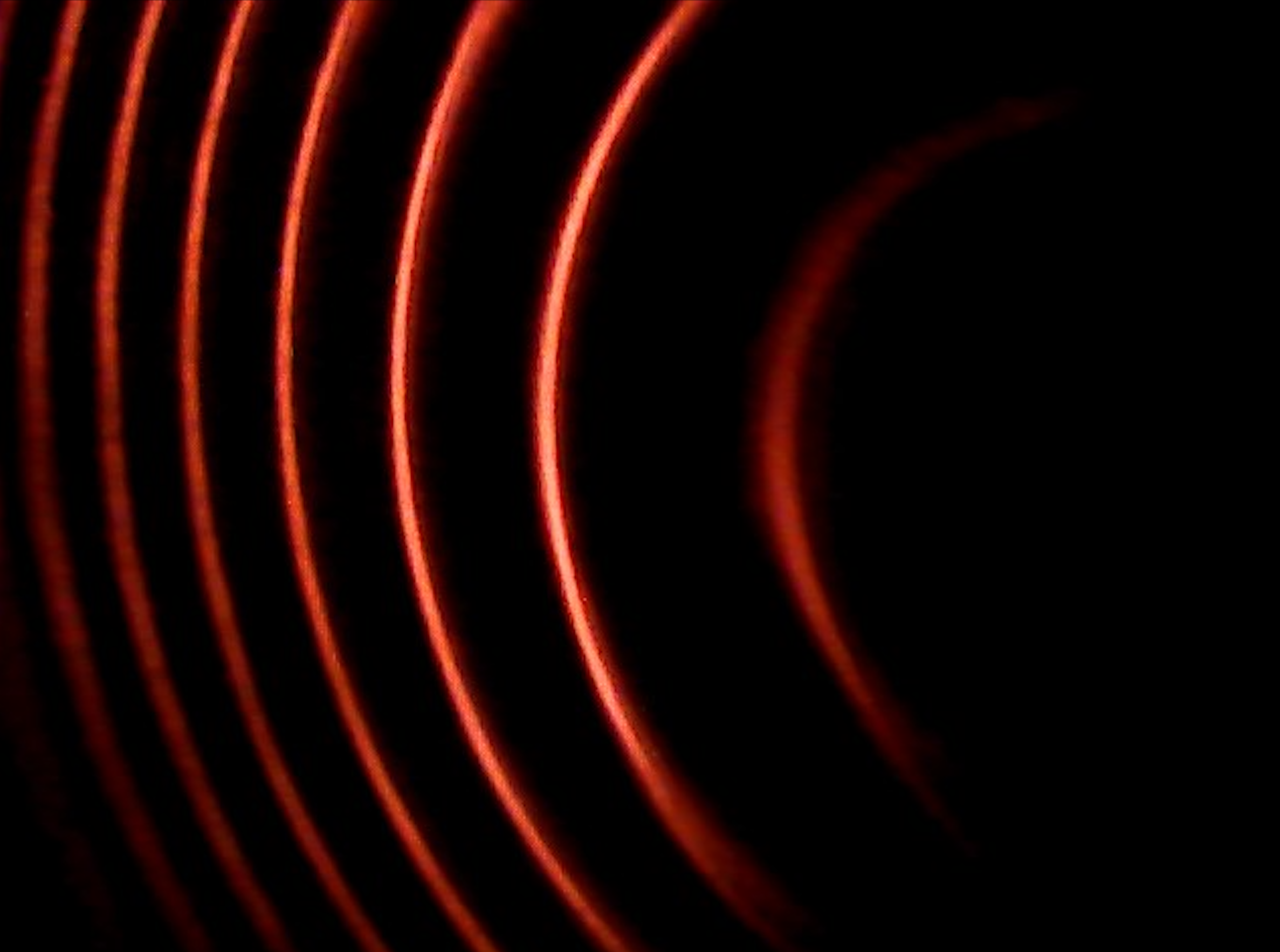}
	\caption{Fringe pattern formed on the camera when OPO signal beam is transmitted through Fabry-Perot interferometer with FSR 7.2 GHz. A single set of rings implies SLM operation of the OPO.}
	\label{etalon}
\end{figure}

For efficient excitation of molecules in an internally cold molecular beam, it is necessary to have SLM operation, stable output wavelength, and narrow linewidth.
As discussed earlier (methods), SLM operation was ensured using an in-line analysis of FPI fringe pattern. 
A single set of concentric rings indicates SLM operation, as shown in Fig. \ref{etalon}. 
Under free-running conditions, we observed that two main factors contribute to the drift in the output wavelength of the OPO.
These are the drift of the pump laser wavelength (typically 0.01 \cm{} over a time scale of 100 seconds) and the change in the OPO cavity length, both of which are independent of each other.
Wavelength stabilization of the idler beam (mid-IR) requires the knowledge of the magnitude and direction of both these drifts (OPO and pump wavelength).
Measurement of the OPO signal wavelength (using the wavemeter), or the PAS signal alone is not sufficient for locking the idler wavelength.
Hence we adopted a different scheme as described below.
Here, a small fraction of the idler beam was sent into the PAS cell such that the PAS signal is proportional to the input power.
Under these conditions the power normalized PAS signal to a good approximation represents the magnitude of the wavelength drift in idler.
Simultaneous measurement of the wavelength of the OPO signal beam using the wavemeter provides the direction of the wavelength drift. 
Using this combined information, the OPO cavity length was adjusted such that the power normalized PAS signal is maximized, ensuring the locking of idler wavelength to the desired spectroscopic transition.
A more detailed description of this scheme is provided in supplementary material.

\subsection{\label{sec:level2} Spectroscopic measurements}   

\begin{figure}
	\centering
	\includegraphics[width=0.48\textwidth]{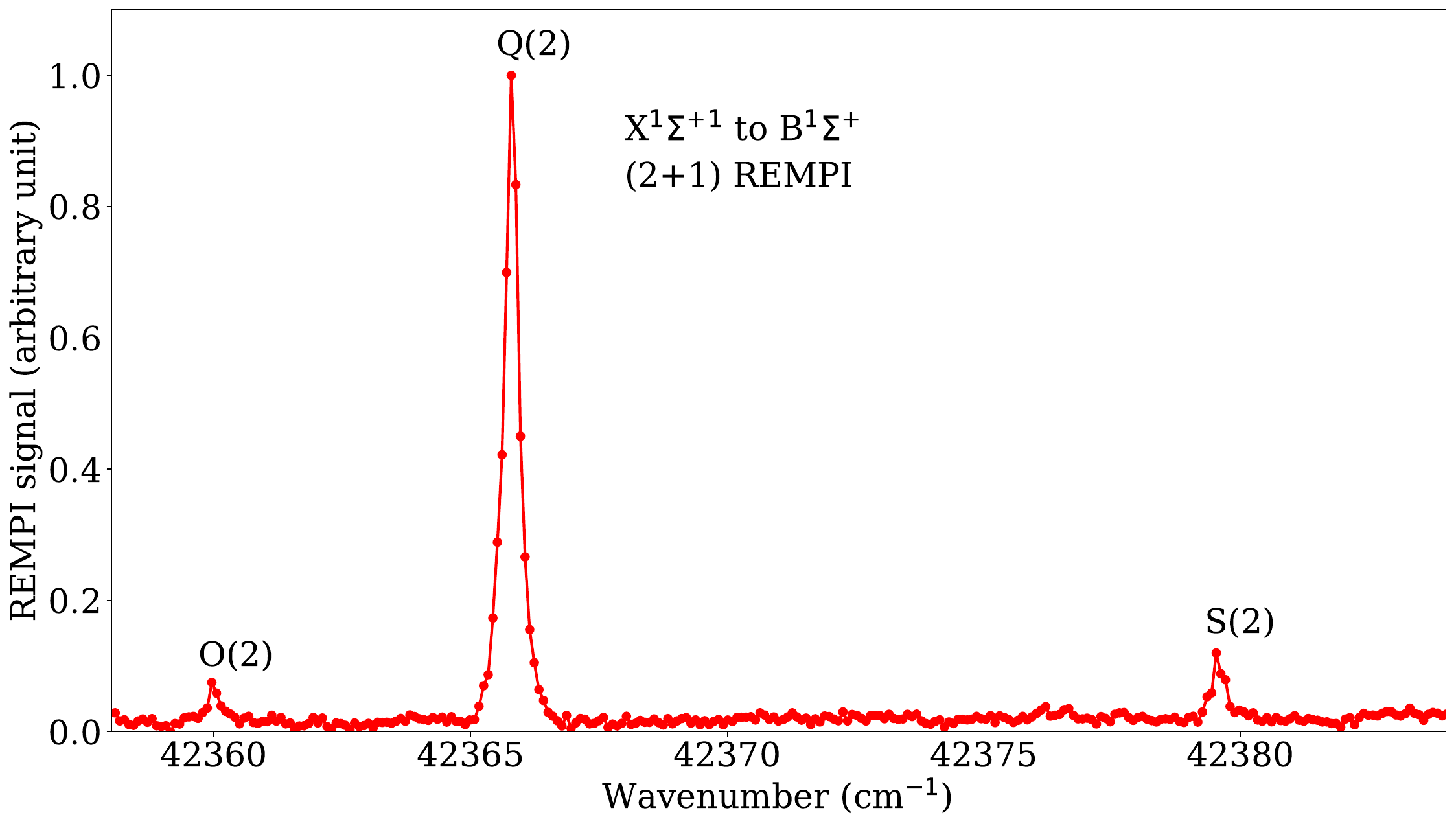}
	\caption{2+1 REMPI spectrum of vibrationally excited CO X$^{1}\Sigma^{+}$(v = 2, J = 2) prepared by IR excitation in the molecular beam.}
	\label{REMPIspectra}
\end{figure}

We have performed several measurements to establish that the ro-vibrational state-selected preparation of molecules in a molecular beam can be done using this setup.
The targeted transitions for demonstrating this were chosen to be CO v = 0 $\rightarrow$ 2 R(0) and R(1), appearing around 4263 \cm{} (2346 nm).
To probe the vibrationally excited CO molecules, we used 2+1 REMPI scheme via  B$^{1}\Sigma^{+}$ state. 
In our experiments, we observed a typical rotational temperature of the order of 10 K for CO (seeded in He) in the molecular beam.
Under these conditions around 40 $\%$ of the CO molecules were present in v = 0, J = 1 state.
Fig. \ref{REMPIspectra} shows the REMPI spectrum of the vibrationally excited CO (v = 2, J = 2)  molecules, which were prepared using IR excitation.
As expected, by scanning the UV wavelength (keeping IR wavelength fixed), three peaks corresponding to the O(2), Q(2), and S(2) transitions were observed (see Fig \ref{REMPIspectra}).
Having established that vibrationally excited CO in a molecular beam can be prepared and detected using this setup, we proceeded further for linewidth characterization of the idler beam.

\begin{figure}
	\centering
	\includegraphics[width=0.46\textwidth]{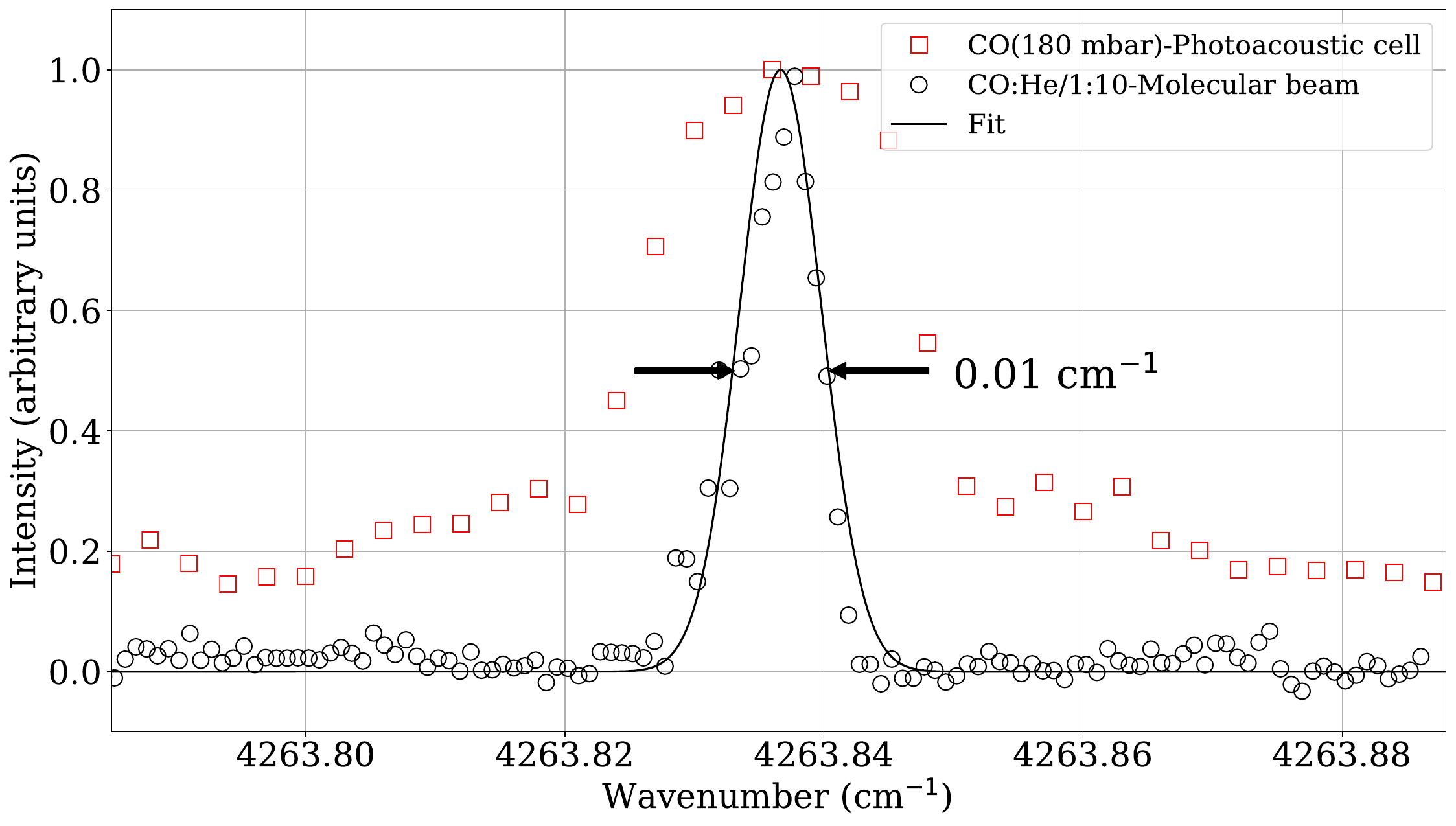}
	\caption{
		The vibrational excitation spectrum of CO v = 0 $\rightarrow$ 2 R(0), measured in a gas cell (180 mbar CO, 300 K), using the PAS method (red squares). Black circles depict the same in a cold molecular beam, measured using IR-UV double resonance. The black curve represents the best fit (Gaussian) resulting in a FWHM of 0.01 \cm{}.
		}
	\label{linewidth}
\end{figure}

For evaluating the linewidth, the idler wavelength was scanned (by scanning the OPO) maintaining SLM conditions, while measuring the REMPI signal of vibrationally excited CO in the molecular beam. 
As an example, a result of one such measurement using the idler beam tuned to R(0) transition is shown in Fig. \ref{linewidth} (black circles).
The black curve shows a fit (Gaussian) resulting in a FWHM of 0.01 \cm{}.
For comparison, data obtained using photoacoustic measurement in a gas cell (180 mbar, 300 K) is also shown (red squares).
Clearly, the observed spectral width is much narrower in the molecular beam. 
Given that the residual Doppler broadening is estimated to be about 5$\times10^{-4}$ \cm{} (based on the angular divergence of our molecular beam), we conclude that the linewidth (FWHM) of the idler beam from our OPO/A system is very close to the observed spectral width of 0.01 \cm{}.


\subsection{\textbf{ Time of flight measurements}}

\begin{figure}
	\centering
	\includegraphics[width=0.45\textwidth]{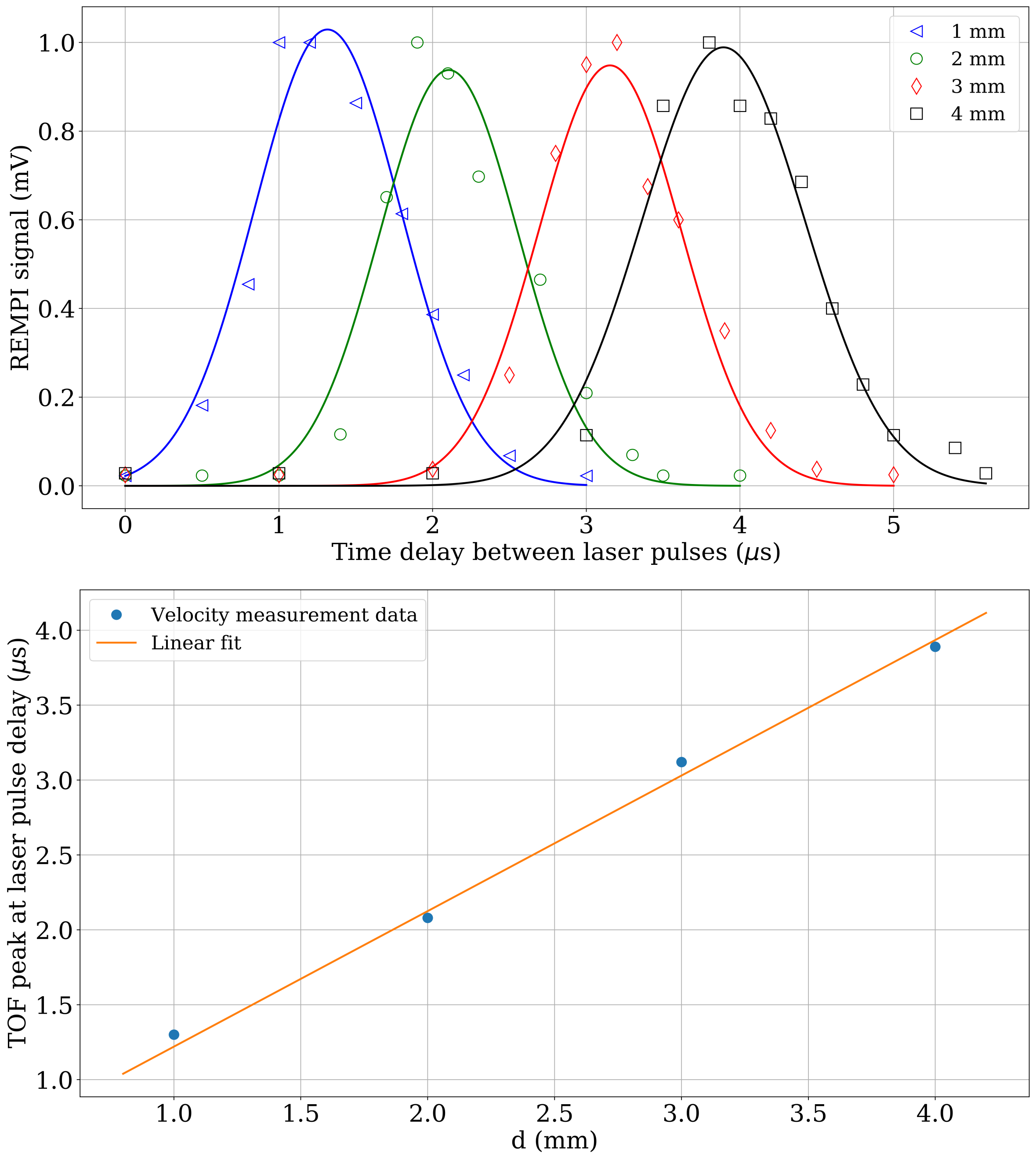}
	\caption{
		 (Upper panel) Time-of-flight measurements carried out by IR-UV double resonance measurement. The IR and UV beams were spatially separated and the resulting arrival time distributions of the vibrationally excited CO are shown along with the best fits (Gaussian, solid lines). 
		 (Lower panel) Peak arrival time distribution plotted as a function of IR-UV separation distance.
		 The orange line depicts the best fit (linear) and its slope was used for determining the mean speed.
	 }
	\label{tof}
\end{figure}

In order to further test the suitability of this system for quantum state-selected scattering experiments, we performed time of flight measurements on the molecular beam to estimate its mean speed \cite{Wodtke_UHV_instrument_2007}.
Here, the CO molecules were excited to a single ro-vibronic state using IR excitation and were subsequently detected downstream in a state-specific manner using REMPI.
The arrival time distributions obtained in this manner were used to estimate the mean velocity and kinetic energy of the CO molecules.
As an example, the results of multiple measurements where IR and UV beams were spatially separated by 1 mm, 2 mm, 3 mm, and 4 mm, respectively, are shown in Fig. 6 (top panel).
The mean arrival time, obtained from the arrival time distributions, as a function of distance is shown in Fig. \ref{tof} (bottom panel).
Based on these results we estimate that the CO molecules to be traveling with the mean speed of 1138 $\pm$ 38 m/s, corresponding to a kinetic energy of 0.18 $\pm$ 0.01 eV.
These results show the potential of our system for carrying out state-selected scattering measurements, using which the translation to rotation, translation to vibration energy transfer, and collision-induced vibrational relaxation processes can be studied \cite{shirhatti2018observation,golibrzuch2014incidence,steinsiek2018work}.

\subsection{\textbf{Wavelength stability estimation}}
In order to quantify the wavelength stability of our system, we chose the R(1) transition of CO (v = 0 $\rightarrow$2).
\begin{figure}[H]
	\centering
	\includegraphics[width=0.45\textwidth]{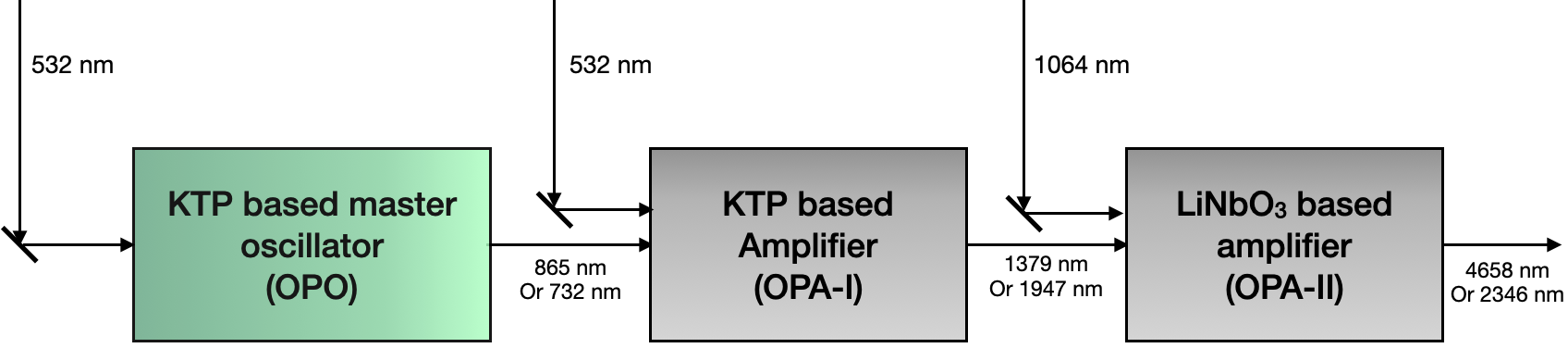}
	\caption{IR wavelength generation using a modified setup. Here, the OPA-II stage consists of a 1064 nm pumped LiNbO$_3$ for difference frequency amplification.
    This scheme allows us to access 1.4 and 4.65 $\mu$m wavelengths. 
 }
	\label{fig:scheme2}
\end{figure}
During the course of our work, we faced problems with some of the anti-reflection coatings on the KTP crystals in the OPA-II stage getting damaged.
As a result, we had to use a slightly modified version of the previously described setup to access 2346 nm (shown in Fig. \ref{fig:scheme2}.
This modification (Scheme 1) involved generating and amplifying a signal beam of 732 nm from OPO, OPA-I stages (using the same KTP crystals), respectively.
Further, a LiNbO$_3$ (cut angle 48$^{\circ}$) based difference frequency generation using a 1064 nm pump was used as OPA-II.

\textbf{Scheme 1} [532 nm (o) = 732 nm (e) + 1947 nm (o)] in OPO [532 nm (o) = 732 nm (e) + 1947 nm (o)] in OPA-I and [1064 nm (e) - 1947 nm (o) = 2346 nm (o)] in OPA-II stage.

\begin{figure}
	\centering
	\includegraphics[width=0.45\textwidth]{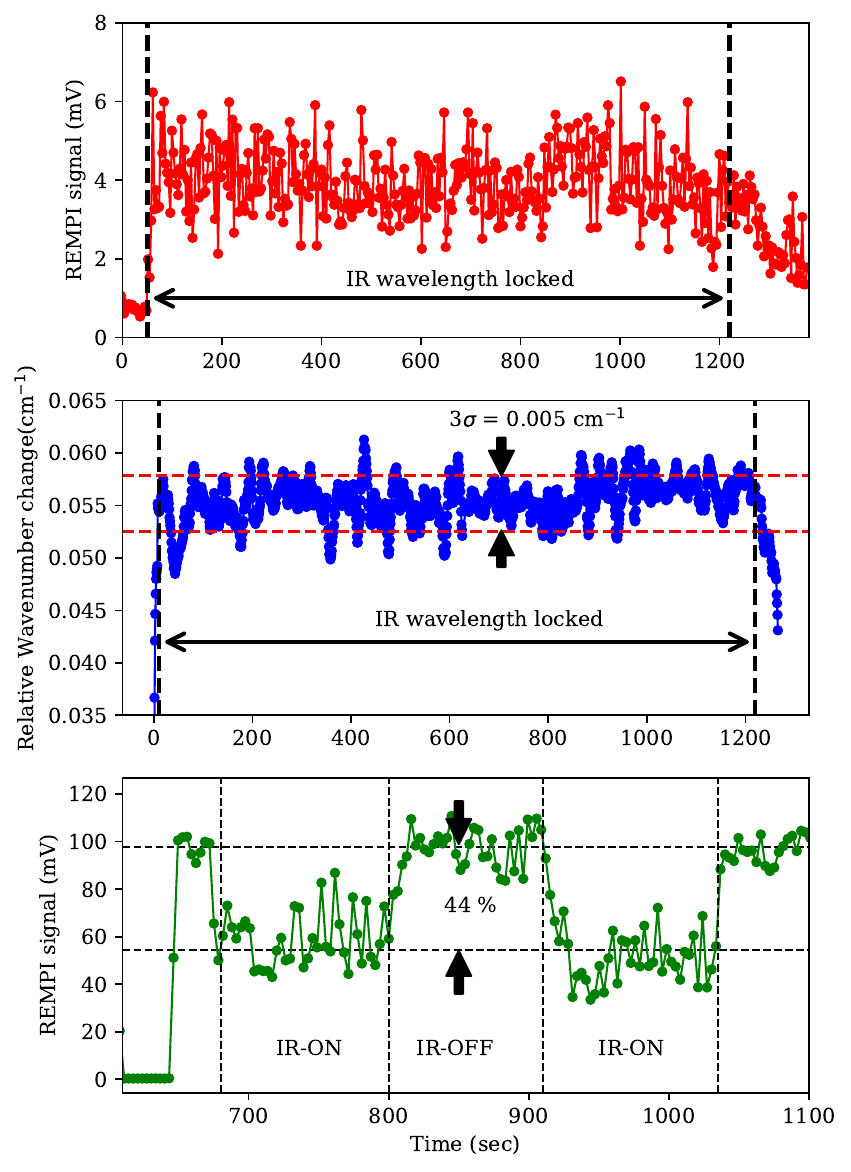}
	\caption{
        (Top panel) REMPI signal vs. time measured for CO v = 2, J = 2 state when IR excitation wavelength is locked. 
		(Middle panel) The idler frequency vs. time was obtained using the calibrated power-normalized PAS signal. These measurements were performed simultaneously (using an independent data acquisition system) along with the REMPI signal shown above.  
        A 3$\sigma$ deviation of 0.005 \cm{} was observed over 1000 sec (red-dashed lines) when the IR wavelength was locked (marked by a double-sided arrow). 
		(Bottom panel) REMPI measurement of CO v = 0, J = 1 state in a molecular beam, with the IR excitation on and off (marked by vertical dashed lines).  
        The upper black dashed line is the average value of the REMPI signal with IR off, and the lower black dashed line represents the same with the IR on. 
        A decrease of 44\% in the ground vibrational state REMPI signal was observed, corresponding to a 70\% population transfer efficiency relative to saturation. 
		}
	\label{depletion}
\end{figure}

This scheme resulted in improved power output at 2346 nm (5-6 mJ/pulse) and also shows the advantage of having such a broadly tunable MOPA-based design. 
Fig. \ref{depletion} illustrates the wavelength stability and overall performance of this setup.
The upper panel shows the REMPI signal as a function of time for the CO prepared in v = 2, J = 2 state using IR-excitation.
This clearly demonstrates the long-term stability of our setup where vibrationally excited CO can be prepared for more than 1000 sec timescale with the IR wavelength locked.
The variation in the observed REMPI signal consists of contributions from fluctuation in the UV and IR power (typically 10\% each, 1$\sigma$, measured over 100 pulses) and IR-frequency fluctuations. 
The middle panel shows the frequency stability of the IR beam, as inferred from the changes in the power-normalized PAS signal, under actively locked conditions (see SI for calibration information).
With the locking on, the IR frequency drift was estimated to remain within 0.005 \cm{} (3$\sigma$), for a duration of more than 1000 sec.  
The bottom panel shows the REMPI signal of the v = 0, J = 1 (ground state) via the S(1) transition with the IR beam on and off.
A clear decrease in the signal, about 44\% on average, due to population depletion in the ground state was observed with the IR beam on. 
Given that the saturation corresponds to a maximum population transfer of 62.5 $\%$ for this transition, our system is able to achieve 70\% population transfer efficiency (relative to saturation).

\begin{figure}
	\centering
	\includegraphics[width=0.46\textwidth]{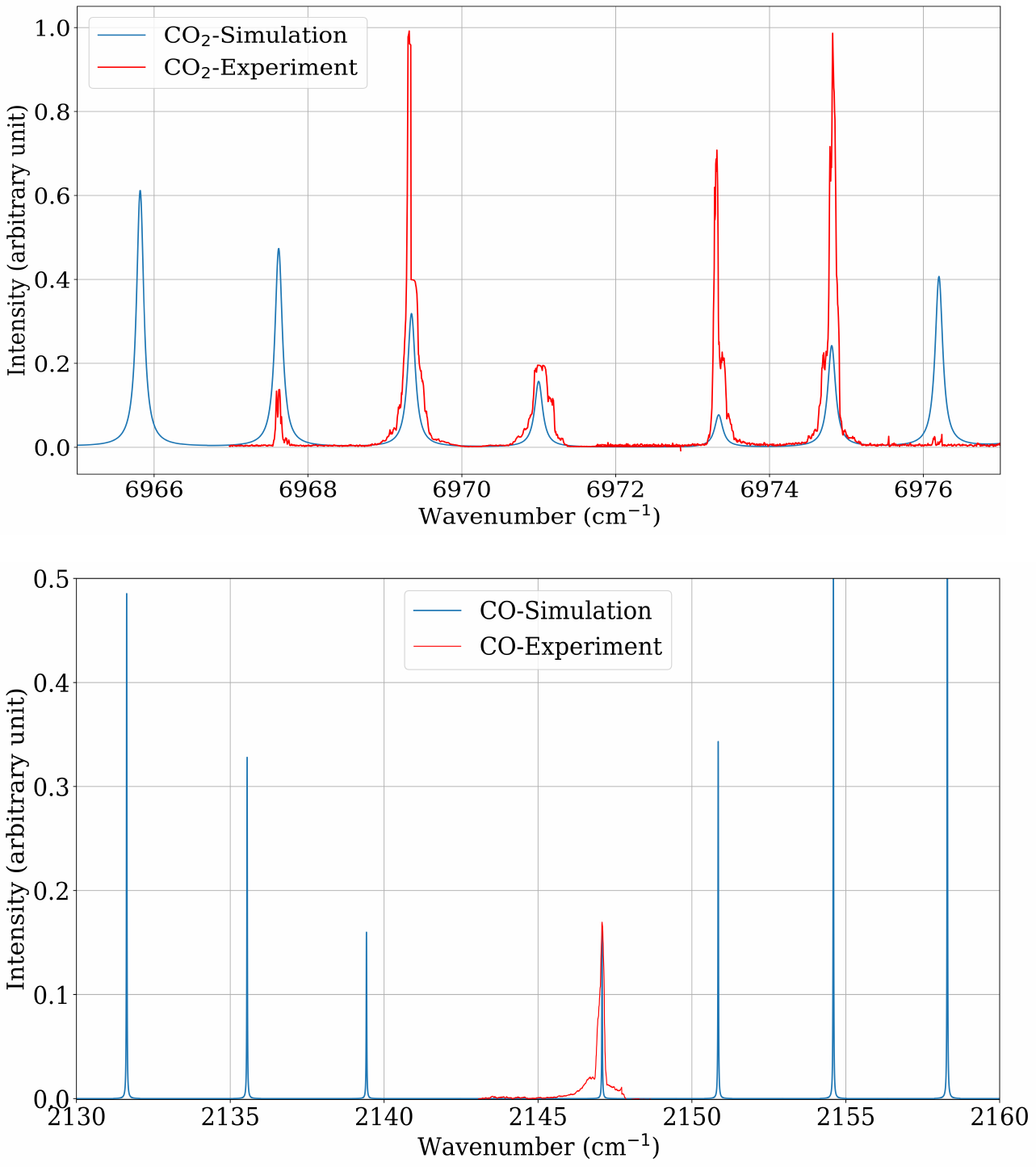}
	\caption{
        (Upper panel) PAS measurement of  CO$_{2}$ in overtone (v = 0 $\rightarrow$ 3 asymmetric stretch) region in a gas cell (500 mbar, 300 K).
		(Lower panel)  PAS measurement of CO (v = 0 $\rightarrow$ 1) in the fundamental region in a gas cell (100 mbar, 300 K). 
		Red curves show the experimental data and blue curves show the simulated spectra using the HITRAN database. In these measurements, SLM feedback was not implemented and the experimental spectra are not corrected for IR power variations (see text).}
	\label{other_wl}
\end{figure}

The wavelength regions accessible using our OPO/A setup can be easily extended with suitable modifications as described below. \\
\textbf{Scheme 2:} [532 nm (o) = 865 nm (e) + 1379 nm(o)] in OPO [532 nm (o) = 865 nm (e) + 1379 nm (o)] in OPA-I and [1064 nm (e) - 1379 nm (o) = 4658 nm (o)] in OPA-II stage. \\
Using a different KTP crystal (cut angle $\theta$/$\phi$ = 64.6/0 degree) and a different LiNbO$_{3}$ crystal (cut angle $\theta$/$\phi$ = 52.3/0 degree) we were able to access the wavelengths in the range of  1.4 $\mu m$  and  4.65 $\mu m$.
As an example, we demonstrate the generation of radiation in 1.4 $\mu m$ region (10 mJ/pulse) for CO$_{2}$, v = 0 $\rightarrow$ 3  asymmetric stretch excitation (Fig. \ref{other_wl} upper panel).
Further using DFG, we access the 4.65 $\mu m$ region, generally considered to be difficult using well-known nonlinear optical materials, for CO v = 0 $\rightarrow$ 1 excitation (Fig. \ref{other_wl} lower panel).
In both these panels, the red curve shows the experimentally measured PAS signal by wavelength scanning the OPO/A system and the blue curves depict the results from simulations using the HITRAN database \cite{Hitran}. 
At a given incidence angle of the pump beam on the KTP crystal, wavelength scan can be performed only within the phase-matching bandwidth. 
In the upper panel, wavelength scan is performed at two different incidence angles, from 6967 \cm{} to 6972 \cm{} and 6972 \cm{} to 6977 \cm{}.  
While scanning the wavelength, the intensity of the radiation also changes, with a peak value at the central wavelength of the phase-matching bandwidth.
The spectra shown here are not corrected for these intensity variations.
These spectral measurements were performed solely to demonstrate the wavelength accessibility of our system and SLM feedback was not implemented here, leading to broad features observed in the spectra.
Using the methods described above, SLM operation can be implemented here as well.
A limitation of this scheme is the large absorption of 4.65 $\mu m$ radiation by the LiNbO$_{3}$ crystal itself (90$\%$). 
This restricts the output energy to 0.1 mJ/pulse (200 mJ/pulse at 1064 nm and 10 mJ/pulse at 1379 nm) making it a very inefficient scheme and having operating parameters very close to the damage threshold of the LiNbO$_3$ crystal.

 \section{\label{sec:level1}Concluding remarks}
The IR-OPO/A system described in this work is versatile and capable enough to meet the stringent requirements of ro-vibrational state-selected preparation of molecules in a cold molecular beam.
The end-to-end solution for building this setup, along with its characterization and wavelength stabilization has been described in detail in this work. 
The most important characteristic of this system is its versatility in terms of wide wavelength tunability.
This design is also promising for generating narrow linewidth radiation in the mid-IR beyond 4 $\mu$m region, especially when combined with nonlinear gain media such as CdSiP$_{2}$ \cite{schunemann2008new,schunemann2009cdsip2} and Orientation-patterned gallium phosphide \cite{kara2017dual}.
Alternatively, this system is promising as a pump source for multipass H$_{2}$ filled Raman cell \cite{hartig1979broadly,rabinowitz1986continuously,antognini2005powerful} for mid-IR wavelength generation over a large range of wavelengths.
In the near future, we plan to explore these possibilities along with applications of our IR-OPO/A setup for state-selected scattering experiments.

\section*{Supplementary Material}
A detailed description of the wavelength locking procedure and related calibration methods.

\section*{Author contributions}
AK contributed to the design, implementation, and testing of the OPO/A setup and the frequency diagnostics and locking schemes with inputs from PRS.
SKS contributed to developing the photoacoustic spectroscopy setup, wavemeter, and the REMPI detection of CO with inputs from AK and PRS.
PRS conceptualized the project.
AK and PRS prepared the manuscript inputs from SKS.
All authors discussed the results and contributed to the manuscript.

\section*{Acknowledgement}
We acknowledge the support of intramural funds at TIFR-Hyderabad provided by the Department of Atomic Energy, Government of India, under Project Identification No. RTI 4007.
We thank Dean Guyer (Laser Vision, USA), for the useful discussions regarding OPO designs and for providing a few components for the master oscillator at the initial stages of our work. 
We also thank Subhashish Dutta Gupta (University of Hyderabad and TIFR-Hyderabad) for useful discussion related to nonlinear optics, Dirk Schwarzer (Max Planck Institute for Multidisciplinary Sciences, G\"{o}ttingen, Germany) for providing LiIO$_{3}$ crystal for DFG stage testing, Dipanjana Saha for preliminary study on OPO designs and Aditi Pradhan for contributing towards testing of photoacoustic spectroscopy setup.
Finally, we thank the institute mechanical workshop members, Rakesh Moodika and Bramha S. Kolpuse for fabricating several custom-designed components used in our setup.

\bibliographystyle{unsrt}
\bibliography{Reference.bib}

\end{document}


\preprint{APS/123-QED}
	
	\title{A versatile and narrow linewidth infra-red radiation source for ro-vibration state selected preparation of molecules in  molecular beams - Supplementary Material}
	
	\author{Avinash Kumar}%
	\affiliation{%
		Tata Institute of Fundamental Research Hyderabad, India
	}%
	\author{Saurabh Kumar singh}%
	\affiliation{%
		Tata Institute of Fundamental Research Hyderabad, India
	}%
	\author{Pranav R. Shirhatti}%
	\email{pranavrs@tifrh.res.in}
	\affiliation{%
		Tata Institute of Fundamental Research Hyderabad, India
	}%

	\date{\today}
	\maketitle
	\section{\label{sec:level1} Wavelength locking setup and calibration}
\begin{figure}
	\centering
	\includegraphics[width=1\textwidth]{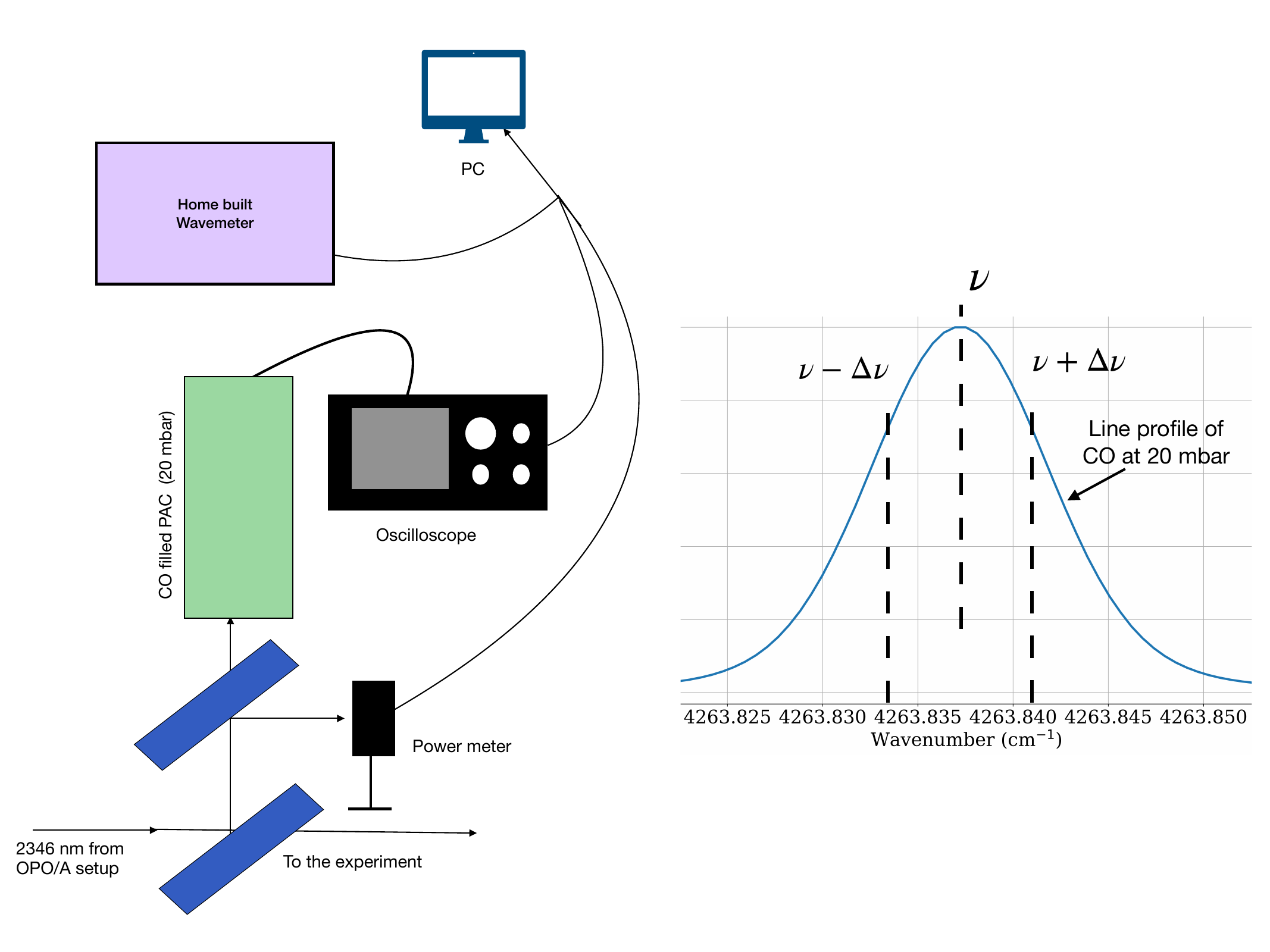}
	\caption{(Left panel) Schematic of the locking setup and (Right panel) line profile of 20 mbar CO at 300 K, generated from HITRAN simulation}
	\label{setup}
\end{figure}

To lock the idler wavelength, we have used a scheme based on photoacoustic spectroscopy (PAS) signal combined with simultaneous wavelength and power measurements.
Fig. \ref{setup} describes the schematic of the locking setup along with wavelength drift analysis plot. 
Here a fraction of the idler beam was sent in to the CO filled (around 20 mbar) absorption cell as input and absorption signal is measured on the oscilloscope.

When wavelength is on resonance, we observe the maximum PAS signal.
If there is a decrease in the PAS signal,
this could be due to power drift of the idler beam, or the wavelength drift or both.
The power drift of the idler depends on the pump laser, which can not be actively controlled. Therefore we needed to eliminate the effect of power drift from the observed PAS signal.
To do this, we used a power meter to measure the power drift and normalize the PAS signal by dividing with measured instantaneous power.
Resultant power normalized PAS (PN-PAS) signal drift now only depends on the wavelength drift.

The wavelength drift could be in positive or negative direction.
 In both cases, signal decreases (see Fig. \ref{setup} right panel).
Therefore it is necessary to know the direction of the wavelength drift also. 
To do this we simultaneously monitor the signal wavelength generated from the OPO stage using the wavemeter.
With all these information combined, and using a LabVIEW based PID algorithm program, we were able to lock the system.

\subsection{\label{sec:level1}Wavelength vs PN-PAS calibration}

For PN-PAS signal measurement, we use only a fraction of the idler beam which is far from saturation, therefore change in PAS signal is directly proportional to the wavelength drift.
To find out proportionality factor of this relation, calibration of the wavelength vs PN-PAS signal is done.

\begin{figure}[h]
	\centering
	\includegraphics[width=1\textwidth]{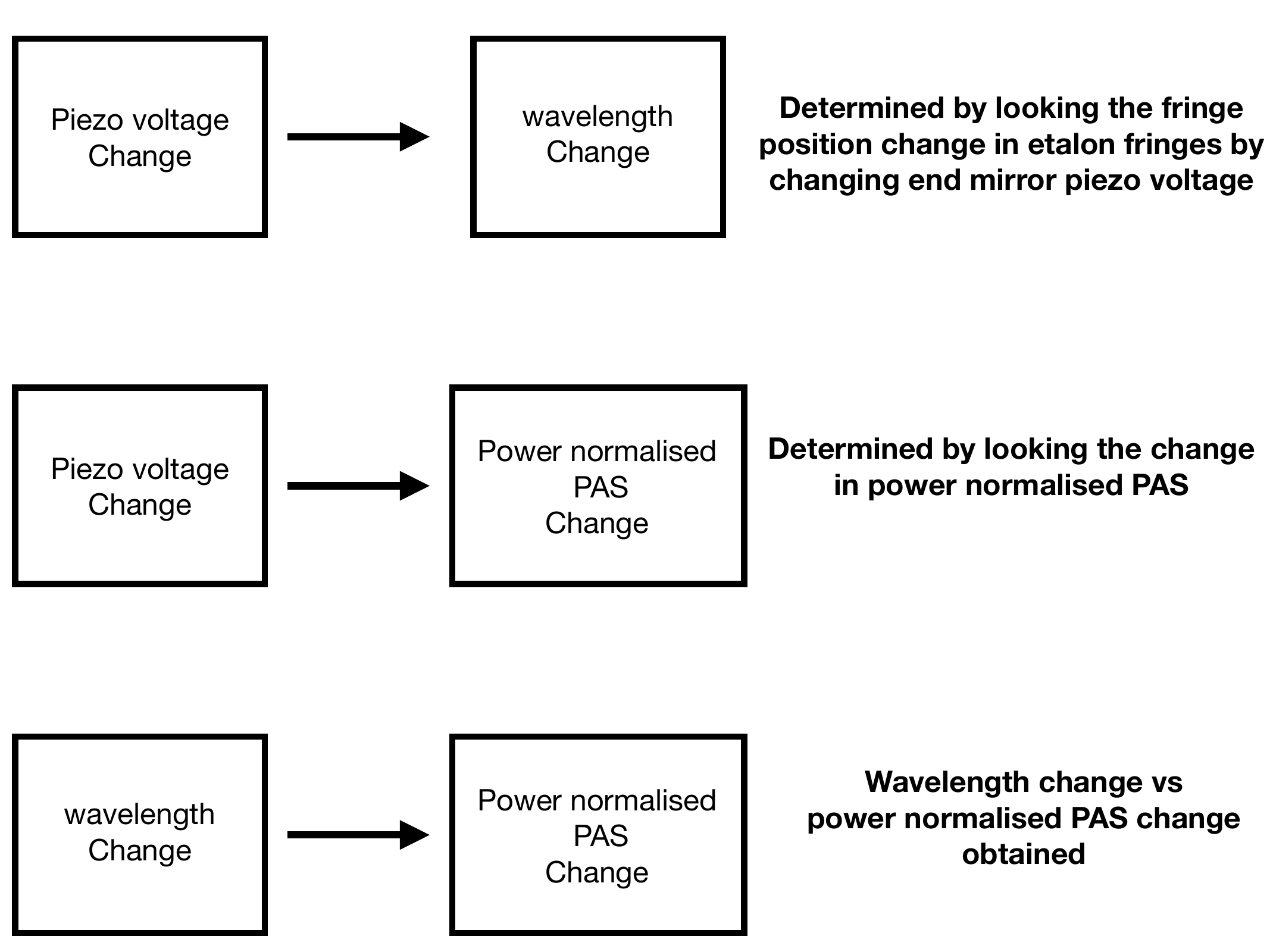}
	\caption{Calibration method flow-chart}
	\label{Calibration}
\end{figure}

We used a Fabry-Perot interferometer (FPI) arrangement (with ability to detect 0.005 cm$^{-1}$ change) to detect wavelength drift. 
Fig. \ref{Calibration} depicts the flow chart for proportionality factor determination between wavelength drift and PN-PAS signal change.


\subsubsection{Piezo voltage vs wavelength calibration}
\begin{figure}[h]
	\centering
	\includegraphics[width=1\textwidth]{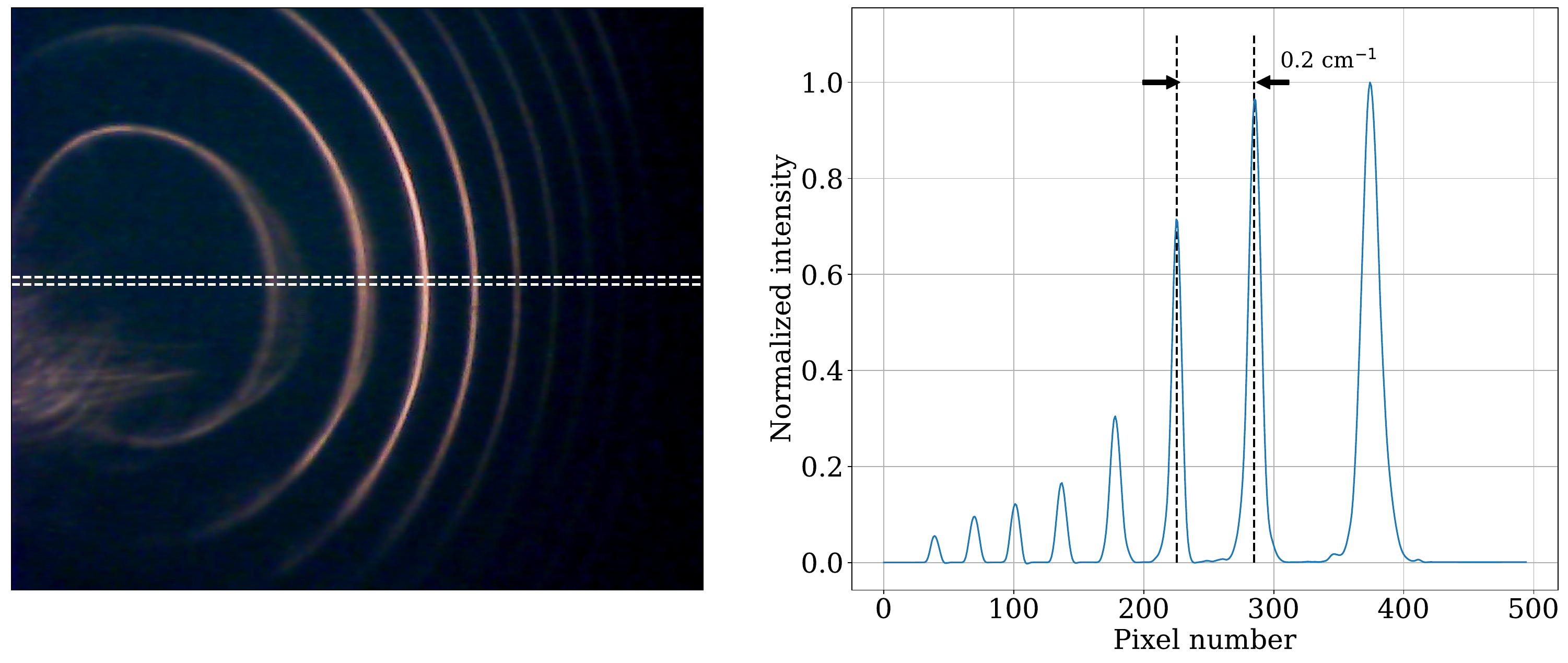}
	\caption{(Left panel) Fringe pattern of the FPI. (Right panel) Intensity distribution of the fringe pattern, in the region within the white horizontal dashed lines.}
	\label{etalon}
	\includegraphics[width=1\textwidth]{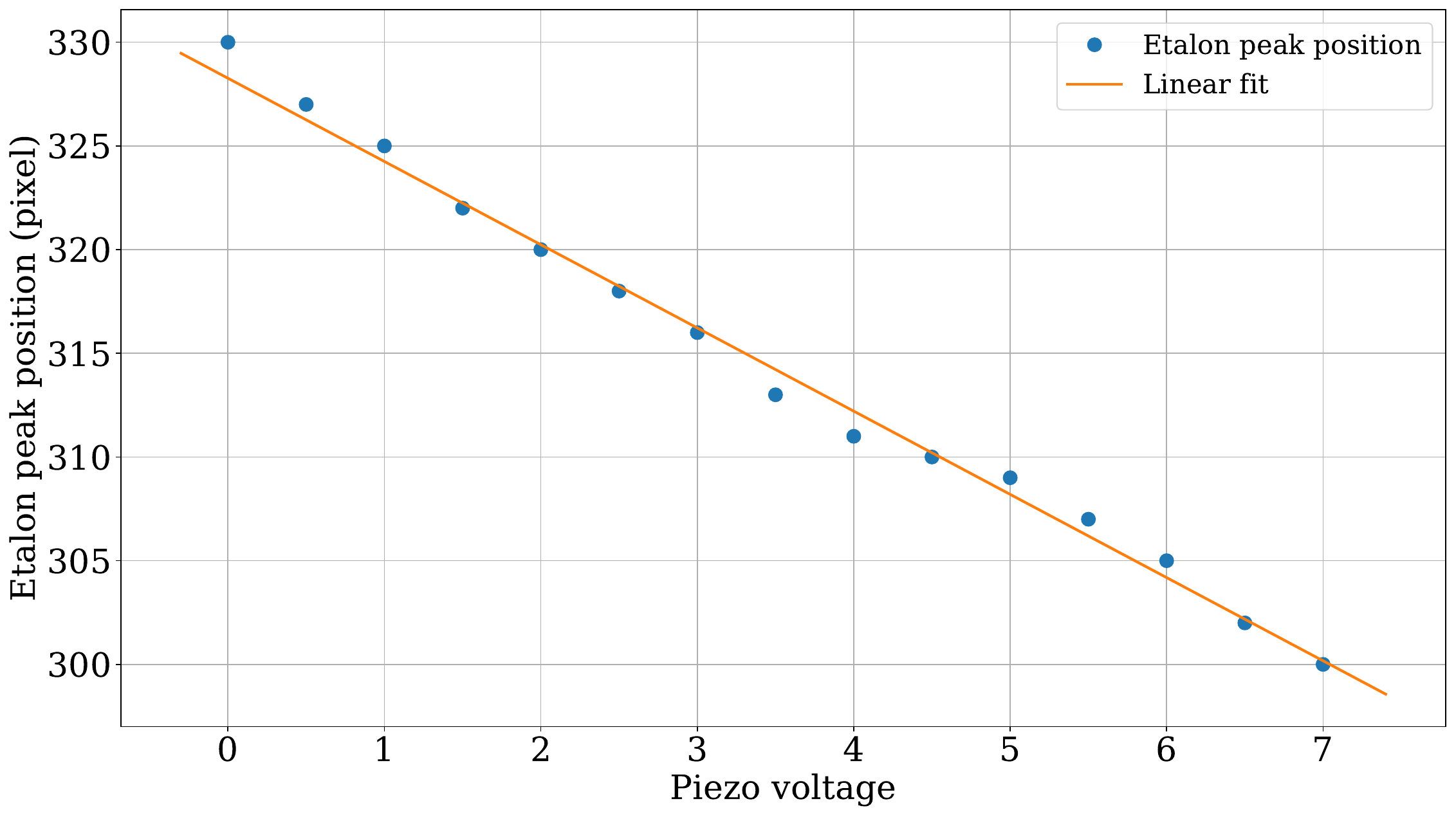}
	\caption{Voltage vs FPI-fringe position plot}
	\label{Voltage}
\end{figure}

In Fig. \ref{etalon}, a fringe pattern of FPI and its intensity distribution along in the region enclosed by the white dashed lines in horizontal direction are shown. 
From the intensity distribution plot (Fig. \ref{etalon}, Right panel), 60 pixels correspond to 0.2 cm$^{-1}$, this implies that, 1 pixel = 0.0033 cm$^{-1}$.

With multiple measurements of voltage and corresponding fringe positions, we estimated the piezo voltage vs wavelength change calibration factor.
Fig. \ref{Voltage} shows the relative peak position of the FPI fringes at different piezo voltages.
The slope of the plot in Fig. \ref{Voltage} shows that a 1 V change in piezo voltage corresponds to 4 pixels shift in fringe position. 
Since 1 pixel change corresponds to 0.0033 cm$^{-1}$ change in wavenumber, we obtain the following relation:  
1 Volt =  4 pixel = 4 $\times$ 0.0033 cm$^{-1}$ = 0.013 cm$^{-1}$

\subsubsection{Voltage vs PN-PAS signal calibration}
Further, for PN-PAS signal vs wavelength calibration, we measured PN-PAS signal as we change the piezo voltage.

\begin{figure}[h]
	\centering
	\includegraphics[width=1\textwidth]{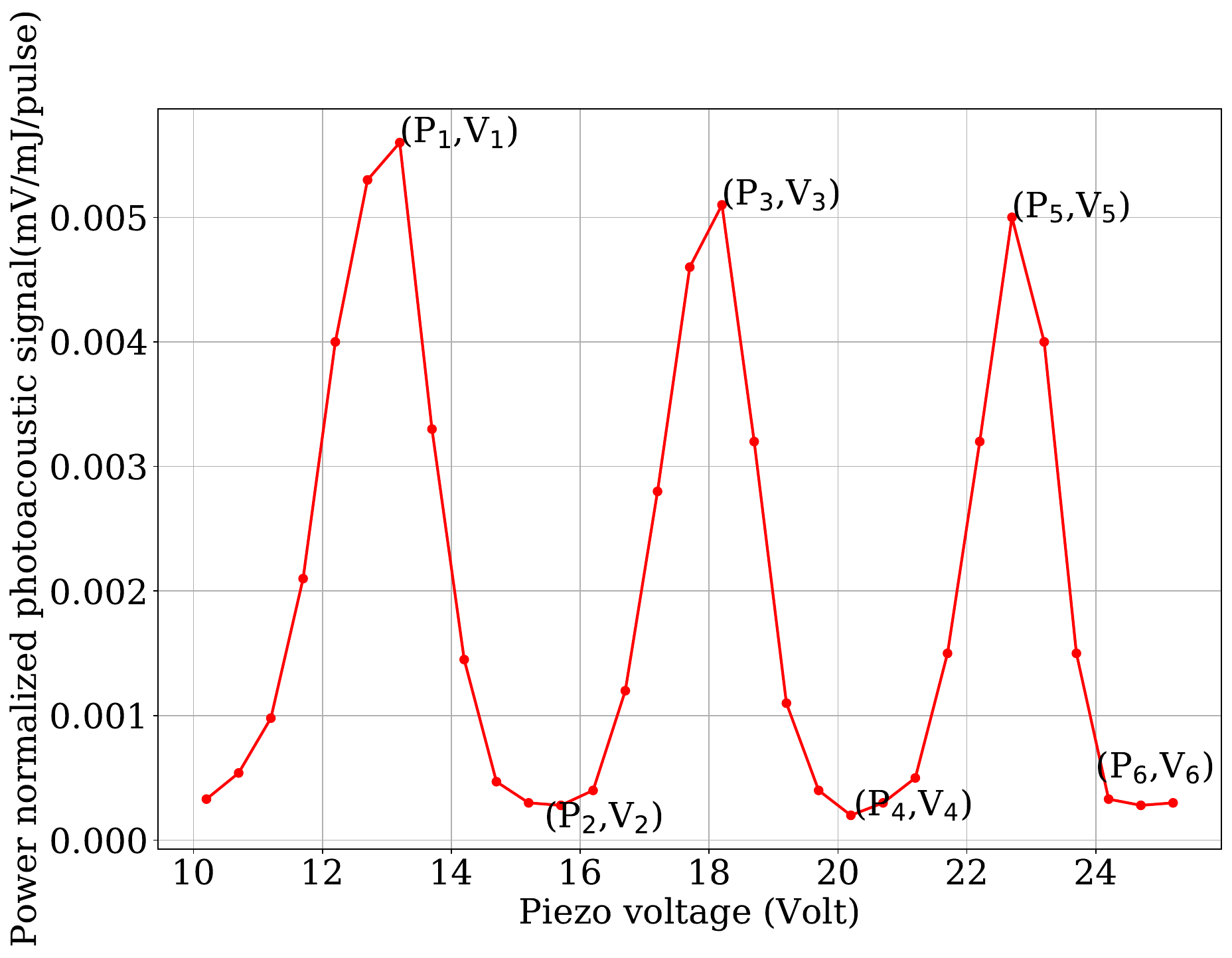}
	\caption{Change in power normalized photoacoustic signal vs piezo voltage.
    The oscillatory behaviour is due to different modes appearing as we scan the cavity length. }
	\label{PNPASvsPiezovoltag}
\end{figure}
%

From Fig. \ref{PNPASvsPiezovoltag}, $\frac{\text{Pi-Pj}}{\text{Vi-Vj}}$ will be the value of the PN-PAS change corresponding to 1 Volt change in piezo voltage. 
\begin{figure}[h]
	\centering
	\includegraphics[width=1\textwidth]{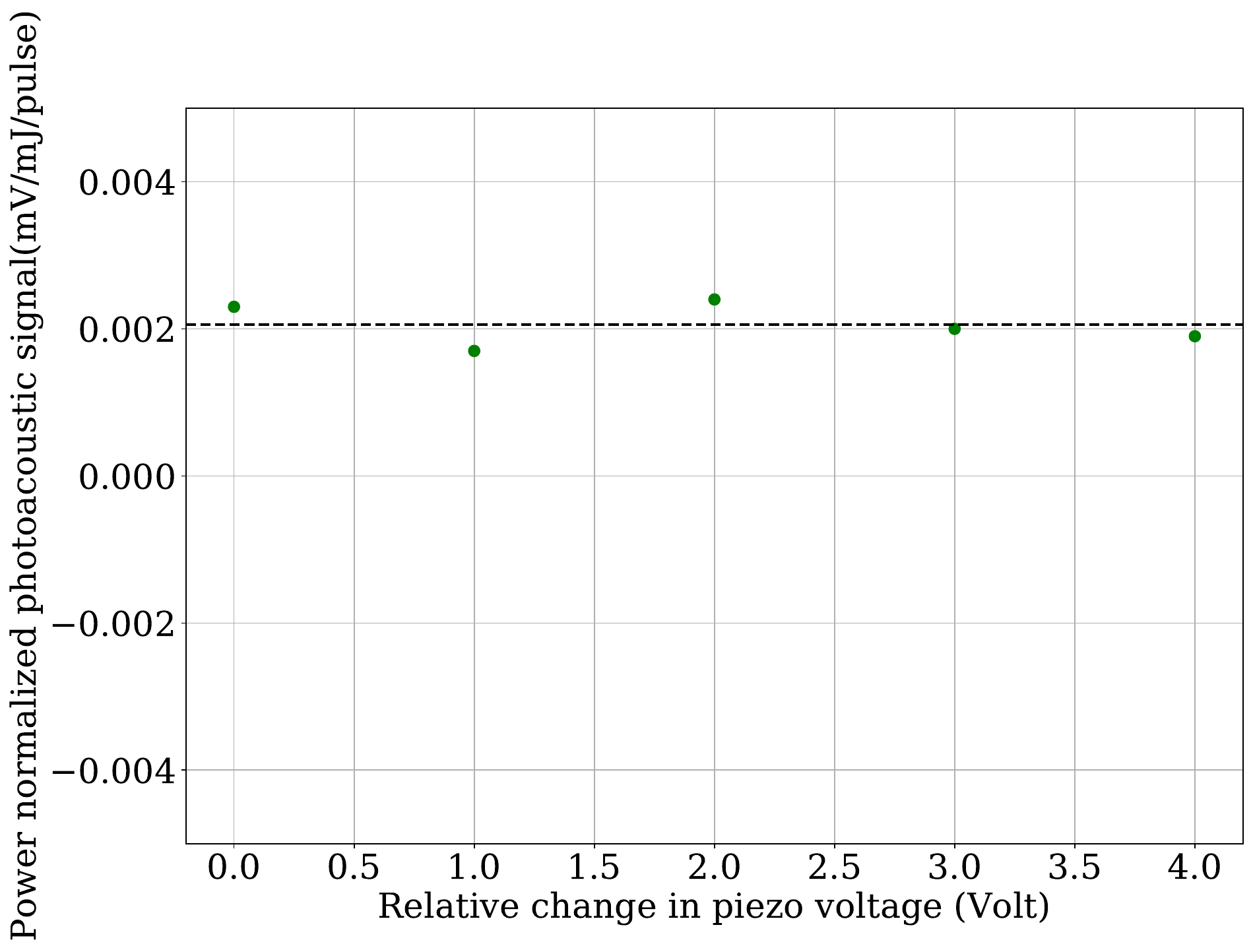}
	\caption{Change in power normalized photoacoustic signal vs piezo voltage. dashed line depicts the mean value and green dots are the calculated values ($\frac{\text{Pi-Pj}}{\text{Vi-Vj}}$) from Fig. \ref{PNPASvsPiezovoltag}.}.
	\label{Relative_voltage}
\end{figure}
Fig. \ref{Relative_voltage} depicts the same using multiple measurements, where we found that with 1 V change in the piezo voltage, the PN-PAS signal changes up to 0.002 unit on an average. 
Therefore, we conclude that 0.002 unit change in PNPAS corresponds to the 0.013 cm$^{-1}$ change in idler wavenumber.